\pdfoutput=1

\documentclass[11pt]{article}

\usepackage{acl}

\usepackage{times}
\usepackage{latexsym}

\usepackage[T1]{fontenc}
\usepackage{wrapfig}
\usepackage{graphicx}
\usepackage{array}
\usepackage{multicol}
\usepackage{multirow}
\newcolumntype{P}[1]{>{\centering\arraybackslash}m{#1}}
\newcolumntype{L}[1]{>{\raggedright\arraybackslash}p{#1}}
\newcolumntype{C}[1]{>{\centering\arraybackslash}S{m{#1}}}
\usepackage[utf8]{inputenc}
\usepackage{amsmath}
\usepackage{amssymb}

\DeclareMathOperator*{\argmin}{arg\,min}
\usepackage{microtype}

\title{UniCoRN: Unified Cognitive Signal ReconstructioN bridging cognitive signals and human language}

\author{
Nuwa Xi, Sendong Zhao\thanks{ \ \ Corresponding author}, Haochun Wang, Chi Liu, Bing Qin and Ting Liu
 \\
Research Center for Social Computing and Information Retrieval, \\Harbin Institute of Technology, China 
\\
\texttt{\{nwxi,sdzhao,hcwang,cliu,bqin,tliu\}@ir.hit.edu.cn}
}

\begin{document}
\maketitle
\begin{abstract}
    Decoding text stimuli from cognitive signals (e.g. fMRI) enhances our understanding of the human language system, paving the way for building versatile Brain-Computer Interface. However, existing studies largely focus on decoding individual word-level fMRI volumes from a restricted vocabulary, which is far too idealized for real-world application. In this paper, we propose fMRI2text, the first open-vocabulary task aiming to bridge fMRI time series and human language. Furthermore, to explore the potential of this new task, we present a baseline solution, UniCoRN: the \textbf{Uni}fied \textbf{Co}gnitive Signal \textbf{R}econstructio\textbf{N} for Brain Decoding.  By reconstructing both individual time points and time series, UniCoRN establishes a robust encoder for cognitive signals (fMRI \& EEG). Leveraging a pre-trained language model as decoder, UniCoRN proves its efficacy in decoding coherent text from fMRI series across various split settings. Our model achieves a 34.77\% BLEU score on fMRI2text, and a 37.04\% BLEU when generalized to EEG-to-text decoding, thereby surpassing the former baseline. Experimental results indicate the feasibility of decoding consecutive fMRI volumes, and the effectiveness of decoding different cognitive signals using a unified structure.
\end{abstract} 
\section{Introduction}

Language serves as a window into the cognitive processes unfolding within our minds, communicating a vast amount of information through its syntax and semantics \citep{pagel2017q}. Advances in cognitive neuroscience have enabled us to directly observe the cognitive processes that underlie language use through the analysis of non-invasive cognitive signals, such as functional Magnetic Resonance Imaging (fMRI) and electroencephalogram (EEG). However, this also poses a challenge in understanding the relationship between these signals and the external stimuli that give rise to them within the mind. Deciphering cognitive signals into human language not only enhances our grasp of the linguistic system, but also facilitates the development of practical brain-computer interfaces (BCIs) by leveraging our comprehension of decoded signals \citep{bci2007,mudgal2020brain}.


Although brain decoding has gained great success from word-level to sentence-level decoding on EEG \citep{panachakel2021survey,wang2022open}, relatively little research has been dedicated to directly generating text, particularly complete sentences, from fMRI volumes. This is largely attributed to the challenges posed by the relatively low temporal resolution of fMRI, which makes it challenging to acquire word-level fMRI frames within a sentence. In this study, we propose fMRI2text, the first open-vocabulary task that decodes fMRI time series into the corresponding texts under naturalistic settings. 

Despite the early efforts in fMRI decoding \citep{mitchell2008predicting,palatucci2009zero,wang2020fine,zou2021towards}, these methods are limited in the ways that they: (1) primarily rely on predefined regions of interest (ROIs) for feature extraction, underutilizing the rich spatial data inherent in full fMRI volumes. This may oversimplify the complex, distributed nature of cognitive processes \citep{ruiz2014real}. (2) do not effectively leverage the sequential information embedded in fMRI time series, missing valuable insights into the dynamics of cognitive processes \citep{du2022fmri}. (3) prioritize the role of the decoder while overlooking the importance of efficient encoding, particularly for high-dimensional signals like fMRI. These limitations extend beyond fMRI decoding and apply to other cognitive signal decoding methods as well. To address these issues and obviate the need for separate, complex pipelines to decode specific cognitive signals, we propose UniCoRN (\textbf{Uni}fied \textbf{Co}gnitive signal \textbf{R}econstructio\textbf{N} for brain decoding), a versatile brain decoding pipeline that can be applied to various types of cognitive signals. 

As a standard encoder-decoder framework, UniCoRN leverages the robust decoding abilities of pre-trained language models. Crucially, it constructs an effective encoder through both snapshot and series reconstructions, harnessing the power of seq2seq models. This allows UniCoRN to analyze individual signal ``snapshots'' (such as a single fMRI volume or an EEG time point) and capture the ``series'' or temporal dependencies among these snapshots, thus maximizing the information extracted from the cognitive signals.



In summary, our contributions are as follows:
\begin{itemize}
    \item We introduce a novel task, designated as fMRI2text, which is the first open-vocabulary task that decodes fMRI time series into human language in a naturalistic context.
    \item We present a baseline solution to further elucidate the potential of fMRI2text and demonstrate that our proposed method is effective across various split settings.
    \item We propose a unified framework UniCoRN (\textbf{Uni}fied \textbf{Co}gnitive signal \textbf{R}econstructio\textbf{N} for brain decoding) to translate cognitive signals into human language, and validate its effectiveness on both EEG and fMRI.
\end{itemize}

\section{Related Work}
\label{related work}
\paragraph{Cognitive Signals}

Cognitive signals represent the dynamic neural activity associated with information processing and cognitive functions, and are crucial in building BCI systems \citep{mudgal2020brain}. These signals are captured at individual time points or as part of a time series, with each data point providing a snapshot of brain activity at a specific point in time. While EcoG is often used in high-performance BCI systems \citep{akbari2019towards,rapeaux2021implantable,metzger2022generalizable}, its semi-invasive nature limits its potential for widespread application in healthy individuals. In non-invasive BCI systems, EEG is most commonly used due to its high temporal resolution and cost-effectiveness, while other techniques such as fMRI have also been employed in recent years \citep{saha2021progress,martinek2021advanced,pitt2022applying}. In spite of its relatively lower temporal resolution, fMRI allows for the mapping of brain-wide responses to linguistic stimuli at a highly detailed spatial resolution of millimeters \citep{vouloumanos2001detection,noppeney2004fmri,binder2009semantic}. This makes fMRI particularly ideal for BCI systems that translate brain signals into text, a process that involves the participation of multiple brain regions \cite{ruiz2014real}.

\paragraph{Brain Decoding}
Recent research has been directed towards resolving the issue of decoding cognitive signals into human language through the introduction of new multi-modal tasks and models. Most recently-proposed tasks in this field focus on aligning cognitive signals with a limited vocabulary up to a thousand for word-level decoding \citep{bhattasali2019localising,affolter2020brain2word,defossez2022decodingeeg} or incorporating them into sentence embeddings for sentence-level decoding using pair-wise classification \citep{pereira2018toward,sun2019towards}. \citeauthor{wang2022open} introduce a novel brain decoding task called EEG-To-Text decoding (EEG2text for short), which achieves sentence-level decoding by converting each word-level EEG signal into corresponding text stimuli using pre-trained language models, thereby extending the problem from a closed vocabulary to an open vocabulary.

\section{Task Definition}
\begin{figure*}[h]
\centering 
\includegraphics[width=0.9\textwidth]{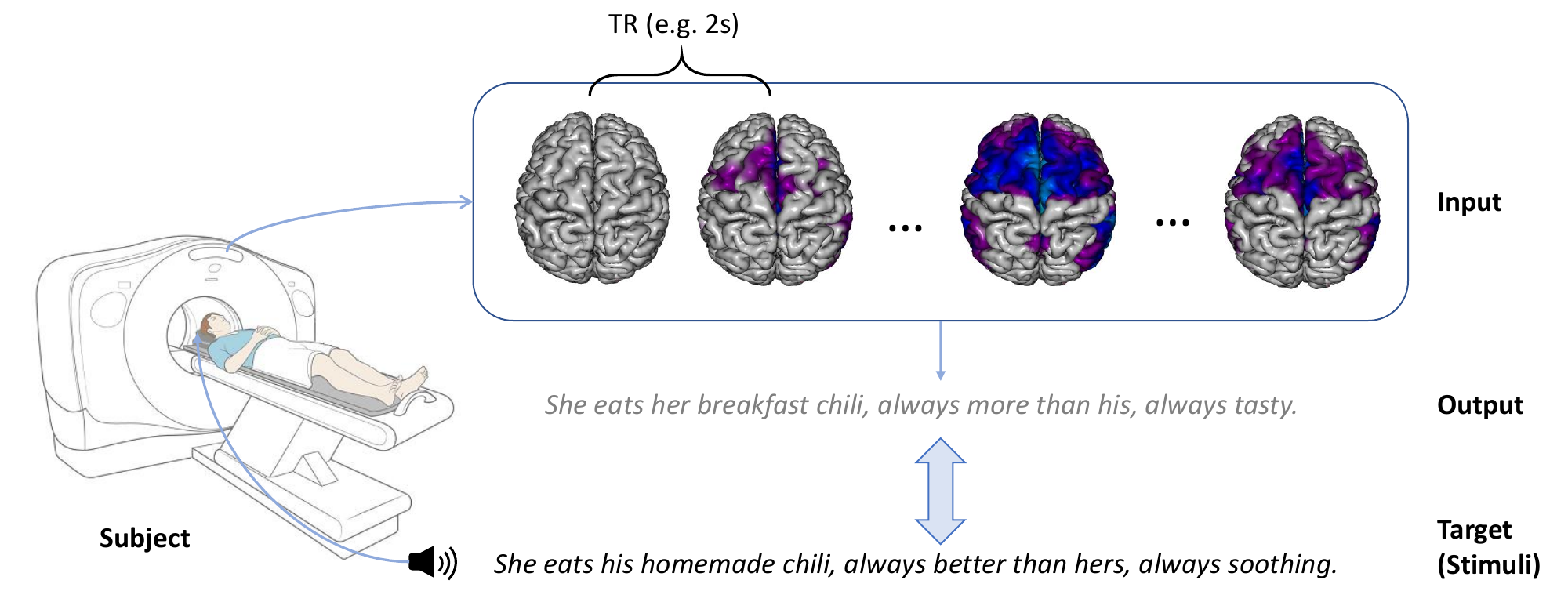} 
\caption{Task definition for fMRI2text} 
\label{task} 
\end{figure*}

\begin{table*}[h]
    \centering
    \begin{tabular*}{0.95\textwidth}{P{0.25\textwidth}|P{0.3\textwidth}P{0.3\textwidth}}
    \hline
    \hline
    
    \textbf{Task Name}&\textbf{Input}&\textbf{Target}\\
    \hline
         Open Vocabulary EEG-To-Text Decoding 
         &a sequence of word-level EEG features $\mathcal{E}:=\{e_1,e_2,...,e_n\}$ 
         &the corresponding text tokens $\mathcal{W}=\{w_1,w_2,...,w_n\}$\\
    \hline
         fMRI-Conditioned Mask-Filling
         &an fMRI image $\mathcal{F}$ and a sentence $\mathcal{W}:=\{w_1,w_2,...,<mask>,...,w_n\}$, where the corresponding word is masked
         &the word masked in sentence $\mathcal{W}$\\
    \hline
        fMRI-Conditioned Text Generation
        &an fMRI image $\mathcal{F}$ and a prefix $\mathcal{W'}:=\{w_1,w_2,...,w_k\}$ 
        &$\mathcal{W}:=\{w_1,...,w_k,...,w_m\}$ where the corresponding word is contained\\
    \hline  
        Open Vocabulary fMRI2text Decoding
        &a fixed-length sequence of $\mathcal{T}$ chronically consistent fMRI $\mathcal{F}:=\{f_1,f_2,...,f_{\mathcal{T}}\}$
        &the correspondent text tokens $\mathcal{W}:=\{w_1,w_2,...w_n\}$\\
    \hline
    \hline
    \end{tabular*}
    \caption{Input and target output for representative brain decoding tasks.}
    \label{task_comparison}
\end{table*}
As shown in Figure \ref{task}, the subject is instructed to read or listen to the text stimuli, while an fMRI volume is acquired every fixed repetition time (TR). Given an fMRI time series of length $\mathcal{T}$, $\mathcal{F}:=\{f_1,f_2,...,f_{\mathcal{T}}\}$, the task is to decode the corresponding text tokens $\mathcal{W}:=\{w_1,w_2,...,w_n\}$ of the stimuli used  during the acquisition of the fMRI volumes from an open vocabulary $\mathcal{V}$. 

As mentioned in Section \ref{related work}, many studies have aimed to link cognitive signals with human language. We summarize three related tasks and compare them to fMRI2text in Table \ref{task_comparison}.

The three representative tasks share a common characteristic of relying on cognitive signals that operate at the word level. However, this approach may not be practical for real-world fMRI applications, as the poor temporal resolution of fMRI necessitates tightly controlled experimental manipulations under these settings \citep{control1,control2}. In contrast, fMRI2text leverages cognitive signals from more naturalistic settings, using text and speech as stimuli in a manner closer to real-world language use \citep{huth2016natural}. Here, each fMRI frame corresponds to a specific timeframe, and is aligned with an undetermined number of tokens rather than a fixed one, better reflecting the variable and dynamic nature of natural language processing.

Another distinct feature that differentiates fMRI2text from prior fMRI-related tasks is its incorporation of multiple sequential frames as input. The inherent low signal-to-noise ratio of fMRI has directed prior studies towards a focus on individual frames. However, this approach  overlooks the valuable temporal information embedded within the interrelations of successive frames, which is particularly crucial when dealing with continuous data streams such as cognitive signals.
\section{Method}
\begin{figure*}[h] 
\centering 
\makebox[\textwidth][c]{%
\includegraphics[width=\textwidth]{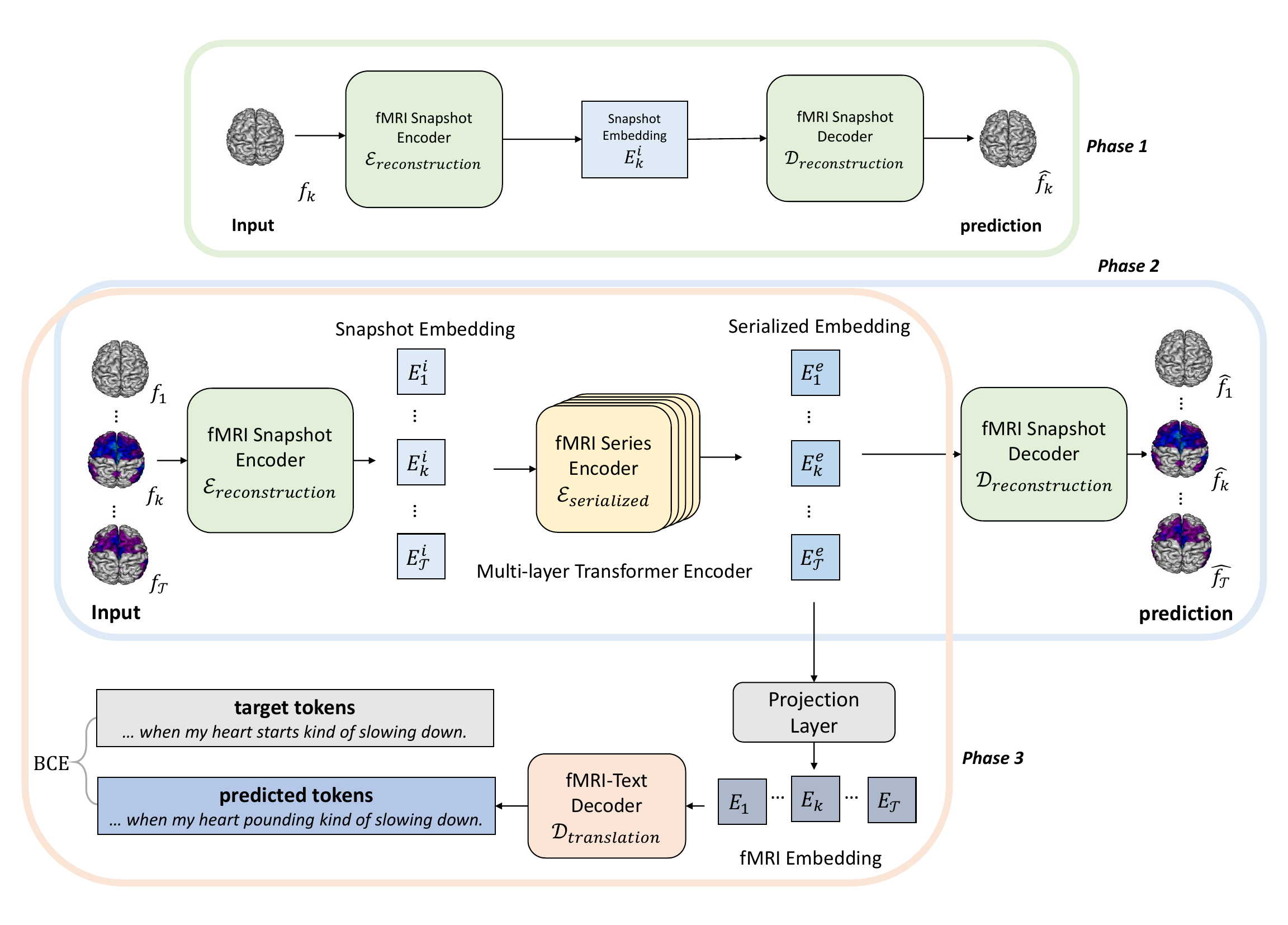}
}
\caption{Illustration of UniCoRN structure. Here we refer to snapshot reconstruction, series reconstruction and cog2text decoding as \emph{Phase 1}, \emph{Phase 2}, \emph{Phase 3} respectively. The snapshot encoder $\mathcal{E}_{r}$ from \emph{Phase 1} (green box) and the series encoder $\mathcal{E}_{s}$ from \emph{Phase 2} (blue box) together yield the fMRI embeddings for fMRI time series, which are then input to the decoder $\mathcal{D}_{t}$ to generate the corresponding text tokens in \emph{Phase 3} (pink box).}
\label{unicorn} 
\end{figure*}

In this section, we introduce the UniCoRN structure and use the fMRI2text task as an explicit demonstration. As shown in Figure \ref{unicorn}, UniCoRN consists of two stages: (1) the cognitive signal reconstruction to train the encoder specifically for cognitive signals, and (2) the cog2text decoding to convert the embeddings of the cognitive signals from the first stage to human language.

\subsection{Cognitive Signal Reconstruction}\label{reconstruction}
The cognitive signal reconstruction consists of two phases, snapshot reconstruction and series reconstruction, aiming to train the encoder of UniCoRN to integrate the individual characteristics of each fMRI volume (intra-volume information), as well as the temporal relationships among volumes in a time series (inter-volume information).

As shown in Figure \ref{unicorn}, during the snapshot reconstruction, each fMRI frame is input into the Snapshot Encoder $\mathcal{E}_r$ ($\mathcal{E}_{reconstruction}$) respectively to obtain the snapshot embedding $E^i$, which will be used later for series reconstruction. In our case, we use a CNN-based model similar to \citet{malkiel2021pre} as the Snapshot Encoder. During this phase, $E^i$ is then fed to the Snapshot Decoder $\mathcal{D}_r$ ($\mathcal{D}_{reconstruction}$) to reconstruct the original fMRI frame $f_k$ (The k-th frame in fMRI time series). Note that $\mathcal{D}_r$ is also CNN-based but simpler than $\mathcal{E}_r$ in structure, to ensure that the reconstruction of fMRI snapshots does not mostly rely on the decoding ability of $\mathcal{D}_r$. We use mean average error (MAE) as the loss function for both phases of cognitive signal reconstruction. \emph{Phase 1} can be formulated as follows:
\begin{equation}
    E^i_k=\mathcal{E}_r(f_k)
\end{equation}
\begin{equation}
    \mathcal{E}_{r}=\argmin_\mathcal{E}\text{MAE}(\mathcal{D}_r(\mathcal{E}_r(f_k)),f_k)
\end{equation}
During \emph{Phase 2}, Series Encoder $\mathcal{E}_s$ ($\mathcal{E}_{serialized}$) takes the snapshot embedding $E^i$ of $\mathcal{T}$ sequential fMRI frames to generate the corresponding serialized embedding $E^e$. We use multi-layer transformer encoder \citep{vaswani2017attention} as $\mathcal{E}_s$ to obtain information in time domain by applying self-attention to fMRI series. Serialized embedding $E^e$ is then input into the same decoder as \emph{Phase 1} for series reconstruction. We continue using $\mathcal{D}_r$ as the decoder to keep minimal effect of decoding process to signal reconstruction, as we will only be using $\mathcal{E}_r$ and $\mathcal{E}_s$ in the next stage. Denote $\{E^e_k,E^e_{k+1},...,E^e_{k+\mathcal{T}-1}\}$ as $E^e_{k\sim\mathcal{T}}$, $\{E^i_k,E^i_{k+1},...,E^i_{k+\mathcal{T}-1}\}$ as $E^i_{k\sim\mathcal{T}}$.
\begin{equation}
    E^e_{k\sim\mathcal{T}}=\mathcal{E}_s(E^i_{k\sim\mathcal{T}})
\end{equation}
\begin{equation}
    \mathcal{E}_{s}=\argmin_\mathcal{E}\text{MAE}(\mathcal{D}_r(\mathcal{E}_s(E^i_{k\sim\mathcal{T}})),E^i_{k\sim\mathcal{T}})
\end{equation}

\subsection{Cog2text Decoding}
The motivation of cognitive signal reconstruction is to get a decent representation of fMRI, which is quite so different from and more difficult than EEG since each fMRI frame has more spatial information as a 3D signal. Similar to \citet{wang2022open}, we use this representation as primary word embeddings for language models, except that these embeddings have been denoised and condensed through reconstruction. The high-level idea here is that we consider each original frame of fMRI as a word-level representation of ``the foreign language spoken by the human brain'', and use the encoder constructed in Section \ref{reconstruction} to obtain the embeddings of this ``language'', which will be then decoded to real human language (English in our case) like traditional machine translation tasks.

Figure \ref{unicorn} gives a detailed demonstration of how fMRI embedding is acquired and how the two stages are concatenated together. After the two phases of cognitive signal reconstruction, the decoder $\mathcal{D}_r$ used in stage one is replaced with the fMRI-Text decoder $\mathcal{D}_t$ ($\mathcal{D}_{translation}$) for text generation. The serialized embeddings $E^e$ are then projected into fMRI embedding $E$ as the final representation of fMRI, which contains both intra-volume information and inter-volume information and will be used as the input for $\mathcal{D}_t$ to convert to texts. Here we use BART \citep{lewis2019bart} as the fMRI-Text decoder $\mathcal{D}_t$ and cross-entropy loss (CE) like most seq2seq tasks as the training target. Denote $\{E_{k},E_{k+1},...,E_{k+\mathcal{T}-1}\}$ as $E_{k\sim\mathcal{T}}$, and the projection layer matrix as $W^P$.
\begin{equation}
    E_{k\sim\mathcal{T}}= E^e_{k\sim\mathcal{T}}W^P
\end{equation}
\begin{equation}
    \mathcal{D}_t=\argmin_\mathcal{D}\text{CE}(\mathcal{D}(E_{k\sim\mathcal{T}}),\mathcal{W})
\end{equation}

\subsection{UniCoRN Structure}

Other than fMRI, UniCoRN is also capable of decoding other cognitive signals into human language. We generalize the same pipeline to EEG2text, without changing the overall structure but only moderately modifying the snapshot encoder $\mathcal{E}_r$ and snapshot decoder $\mathcal{D}_r$ due to the difference in spatial structure between EEG and fMRI. The detailed illustration is provided in Appendix \ref{appendix-eeg}.

\section{Experiments}

\subsection{Dataset}\label{dataset}
The ``Narratives'' dataset \citep{nastase2021narratives} encompasses a range of fMRI data from individuals who were engaged in listening to spoken stories in the real-world setting. Given that various fMRI machines produce frames of different sizes, and considering the ``Narratives'' dataset comprises data from multiple machines, we focus solely on data with dimensions of $64\times64\times27$ voxels. The detailed information of the ``Narratives'' dataset we used in this paper is provided in Appendix \ref{appendix-dataset}. 

Most cognitive signals require pre-processing before putting into use. For fMRI, We follow the same pre-processing procedure as provided in \citet{nastase2021narratives}. As for EEG, we use the same waves as in \citet{wang2022open} for comparison.
\begin{table}[h]
    \renewcommand{\arraystretch}{1.2}
    \resizebox{\columnwidth}{!}{\begin{tabular}{c|c}
    \hline
    \hline
    \textbf{Split Method}&\textbf{Test Set}\\
    \hline
    random&

    $\{F^{ij}_{k\sim\mathcal{T}}|F^{ij}_{k\sim\mathcal{T}}\not\in \mathbb{F}_{Tr}\}$\\
    \hline
    random time&
    $\{F^{ij}_{k\sim\mathcal{T}}|\forall j, k \notin T^{j}_{Tr}\}$
    \\
    \hline
    consecutive time&
    $\{F^{ij}_{k\sim\mathcal{T}}|\forall j, \forall t \in T^{j}_{Tr},t<k\}$
    \\
    \hline
    by stimuli&
    $\{F^{ij}_{k\sim\mathcal{T}}|j \not\in \mathcal{C}_{Tr}\}$
    \\
    \hline
    by subject&
    $\{F^{ij}_{k\sim\mathcal{T}}|i \not\in \mathcal{S}_{Tr}\}$
    \\
    \hline
    \hline
    \end{tabular}}
    \caption{Splitting method for fMRI2text. Detailed notations are further explained in Appendix \ref{appendix-notation}.}
    \label{split_method}
\end{table}

Given that the ``Narratives'' dataset does not offer any pre-determined splits and the appropriate method for splitting fMRI data for this task is a matter of debate, we conduct experiments utilizing a variety of different split configurations.

Denote all subjects as $\mathcal{S}:=\{S_1,S_2,...,S_n\}$, all stimuli as $\mathcal{C}:=\{C_1,C_2,...,C_m\}$, where $n$ and $m$ stands for the total number of subjects and stimuli respectively. Note that the total number of stimuli given to individual subjects may vary. The fMRI series of subject $S_i$ receiving stimuli $C_j$ is represented as $\mathcal{F}^{ij}:=\{f^{ij}_1,f^{ij}_2,...,f^{ij}_{T_j}\}$. $T_j$ here represents the total number of fMRI frames of stimuli $j$. For briefings, we use $F^{ij}_{k\sim\mathcal{T}}$ to represent the fMRI series of length $\mathcal{T}$ starting at the $k$ th frame $\{f^{ij}_k,f^{ij}_{k+1},...,f^{ij}_{k+\mathcal{T}-1}\}$. Different split methods are formulated in detail in Table \ref{split_method}.

As for EEG2text, We use ZuCo1.0 datasets \citep{hollenstein2018zuco}, which comprises EEG recordings obtained from natural reading tasks, including both Normal Reading (NR) and Task-Specific Reading (TSR). The reading materials utilized for these tasks were sourced from movie reviews \citep{socher2013recursive} and Wikipedia articles. The ZuCo1.0 dataset comprises a total of 1,107 unique sentences across 12 subjects, yielding a total of 10,258 samples. Given the limited number of training samples, we utilize a split method similar to the $random$ method described above.

\subsection{Implementation}

\begin{table*}[h]
  \centering
  \makebox[\textwidth][c]{\begin{tabular*}{\textwidth}{P{0.16\textwidth}|P{0.1\textwidth}P{0.1\textwidth}P{0.1\textwidth}P{0.1\textwidth}|P{0.08\textwidth}P{0.08\textwidth}P{0.08\textwidth}}
    \hline
    \hline
    \multirow{2}{*}{\textbf{Method}} &
      \multicolumn{4}{c|}{\textbf{BLEU-N (\%)}} &
      \multicolumn{3}{c}{\textbf{ROUGE-1 (\%)}} \\
      &\textbf{BLEU-1}&\textbf{BLEU-2}&\textbf{BLEU-3}&\textbf{BLEU-4}&\textbf{F}&\textbf{P}&\textbf{R}\\
    \hline
    random &65.64 	&\textbf{52.51} 	&\textbf{44.96} 	&\textbf{39.74} &	60.74 &	63.63 	&58.44
\\
    \hline
    random time&62.90 &	49.00 &	40.59 &	34.77 &	59.52 &	62.65 	&56.91
\\
    \hline
    consecutive time&28.21 &	9.23 &	4.27 	&1.83 &	21.88 	&25.84 &	19.12 
\\
    \hline
    by stimuli&	26.29 &	6.66 	&2.26 	&0.53 &	23.72 &	30.74 	&19.40 
\\
    \hline
    by subject&	\textbf{66.10} &	52.32 	&43.78 &	37.78 &	\textbf{62.68} 	&\textbf{66.06} &	\textbf{59.88} 
\\
    \hline
    \hline
  \end{tabular*}}
  \caption{Results of UniCoRN for fMRI2text on different split settings.}
  \label{split_result}
\end{table*}
\begin{table*}[h]
  \centering
  \makebox[\textwidth][c]{\begin{tabular*}{\textwidth}{P{0.16\textwidth}|P{0.1\textwidth}P{0.1\textwidth}P{0.1\textwidth}P{0.1\textwidth}|P{0.08\textwidth}P{0.08\textwidth}P{0.08\textwidth}}
    \hline
    \hline
    \multirow{2}{*}{\textbf{$\mathcal{T}$}} &
      \multicolumn{4}{c|}{\textbf{BLEU-N (\%)}} &
      \multicolumn{3}{c}{\textbf{ROUGE-1 (\%)}} \\
      &\textbf{BLEU-1}&\textbf{BLEU-2}&\textbf{BLEU-3}&\textbf{BLEU-4}&\textbf{F}&\textbf{P}&\textbf{R}\\
    \hline
    1	&39.16 &	9.62 	&3.47 	&1.09&	11.00 &	12.74 	&10.38 
\\
    \hline
    3	&25.17 &	9.89 	&5.05 	&2.75 	&19.46 &	17.05 	&23.15 
\\
    \hline
    5	&44.78 &	24.95 &	15.75 	&10.58 &	36.49 	&39.90 	&33.95 
\\
    \hline
    8	&49.66 &	30.71 &	21.10 &	15.44 	&43.75& 	48.14 &	40.38 
\\
    \hline
    10&\textbf{62.90} &	\textbf{49.00} &	\textbf{40.59} &	\textbf{34.77} &	\textbf{59.52} &	62.65 	&\textbf{56.91}
\\
    \hline
    12	&62.02 	&47.35 	&38.77& 	33.04 &	59.02 	&\textbf{63.09} 	&55.65 
\\
    \hline
    14&	58.58 &	42.27 	&33.07 &	27.09 &	54.78 &	59.39 	&51.10 
\\
    \hline
    16&51.14 &	32.58 &	22.87 &	17.17 &	46.45 &	51.47 	&42.53 
\\
    \hline
    \hline
  \end{tabular*}}
  \caption{Results of UniCoRN for fMRI2text on different series length $\mathcal{T}$.}
  \label{len}
\end{table*}
\begin{table*}[h]
    \centering
    \makebox[\textwidth][c]{\begin{tabular*}{\textwidth}{P{0.18\textwidth}|P{0.05\textwidth}|l}
    \hline
    \hline
    \textbf{Split Method}
    &\textbf{$\mathcal{T}$}
    & \multicolumn{1}{>{\centering\arraybackslash}m{0.6\textwidth}}{\textbf{Results}}
    \\
    \hline

    \multirow{2}{*}{consecutive time}
    &\multirow{2}{*}{10}
    &T: the policeman, um, he \textbf{doesn’t} even \textbf{say} anything to \textbf{Sherlock}...
    \\
    & &  P: and first, the, she just \textbf{doesn't} \emph{talk} though \textbf{Sherlock}...
    \\
    \hline
    \multirow{2}{*}{by stimuli}
     &\multirow{2}{*}{10}
     &T: I think it's \textbf{some sort of} mass hyp\textbf{nosis} or something...  
    \\
     & & P: and you \emph{a sort of} the You\textbf{nosis} session something...
    \\
    \hline
    \multirow{2}{*}{random time}&
    \multirow{2}{*}{1}
    &T: He woke up early the \textbf{next} \textbf{morning}
    \\
     & & P: I's up and \textbf{morning} \emph{other day}
    \\
    \hline
    \multirow{2}{*}{random time}
    &\multirow{2}{*}{3}
    &  T: she put her \textbf{arm} through mine and squeezed it a little bit. 
    \\
     & & P: I says her \emph{shoulder} through mine and I it a little bit
    \\
    \hline
    \multirow{2}{*}{random time}
    & \multirow{2}{*}{5}
    &  T: Um, it was an \textbf{extremely Darwinian moment} for me, uh, \textbf{because}...
    \\
     & & P: I and, like \emph{best} \textbf{Darwinian moment} for me, and, \emph{for}...
    \\
    \hline
    \hline
    \end{tabular*}}
    \caption{Case Analysis for fMRI2text. The target sentence is denoted as T, and the predicted sentence is represented by P. Text fragments in the target sentence to be compared are in \textbf{bold} font. Exact matches between the target and predicted sentences are indicated in \textbf{bold}, while semantic similarity is shown in \emph{italic} font.}
    \label{case_study_fMRI}
\end{table*}
\begin{table*}[h]
    \renewcommand{\arraystretch}{1.2}
    \centering
    \makebox[\textwidth][c]{\begin{tabular*}{\textwidth}{P{0.03\textwidth}|L{0.92\textwidth}}
    \hline
    \hline
    \multirow{3}{*}{(1)}
    &T: Stephen Rea, \textbf{Aidan Quinn}, and Alan Bates play Desmond's legal \textbf{eagles}... \\
    \hline
    &P: He Hara, \textbf{Aidan Quinn}, and Alan Bates play Desmond's legal \textbf{eagles}... \\
    \hline
    &B: He Baldwina, \emph{Longan} \underline{shows}, and Alan Lloyd play Hannibal's legal \underline{eternally}... \\
    \hline
    \multirow{3}{*}{(2)}
    &T: the \textbf{sight} of this grandiloquent quartet \textbf{lolling} in pretty Irish settings is a \textbf{pleasant} enough \textbf{thing} \\
    \hline
    &P: the \textbf{sight} of this grandiloquent Shet \textbf{lolling} in pretty Irish American is a \underline{lot} enough \textbf{thing} \\
    \hline
    &B: the \underline{real} of this this asquent Shet \underline{filmolling's} grand much American is a \underline{talented} enough \emph{film} \\
    \hline
    \hline
    \end{tabular*}}
    \caption{Case Analysis for EEG2text. T, P, B denote the target sentence, UniCoRN predictions, and baseline predictions, respectively. Text fragments in \textbf{bold} represent the compared portions in the target sentence. \textbf{Bold} highlights or \underline{underlines} indicate representative matches/mismatches, while \emph{italicization} signifies semantic similarity.}
    \label{case_study_EEG}
\end{table*}

\begin{table*}[h]
  \centering
  \makebox[\textwidth][c]{\begin{tabular*}{\textwidth}{P{0.16\textwidth}|P{0.1\textwidth}P{0.1\textwidth}P{0.1\textwidth}P{0.1\textwidth}|P{0.08\textwidth}P{0.08\textwidth}P{0.08\textwidth}}
    \hline
    \hline
    \multirow{2}{*}{\textbf{Method}} &
      \multicolumn{4}{c|}{\textbf{BLEU-N (\%)}} &
      \multicolumn{3}{c}{\textbf{ROUGE-1 (\%)}} \\
      &\textbf{BLEU-1}&\textbf{BLEU-2}&\textbf{BLEU-3}&\textbf{BLEU-4}&\textbf{F}&\textbf{P}&\textbf{R}\\
    \hline
    UniCoRN&	57.68 &	47.93 	&41.73 &	\textbf{37.04} &	\textbf{64.39} &	60.37 	&\textbf{70.00}
\\
    \hline
    w/o p1&	\textbf{59.63} &	\textbf{48.90} 	&\textbf{41.87} 	&36.51 &	62.40 &	59.92 	&66.25 
\\
    \hline
    w/o p2 & 48.51 &	37.15 	&30.25 &	25.28 &	52.49 	&47.48 	&60.94 
\\
    \hline
    w/o p1p2& 57.78 	&46.40 &	39.10 &	33.69 	&62.42 &	\textbf{61.01} 	&64.44 
\\
    \hline
    baseline&	54.02 	&44.93 	&39.09 &	34.65 &	58.78 &	52.75 	&67.87 
\\
    \hline
    \hline
  \end{tabular*}}
  \caption{Results of EEG2text ablation study. \emph{p1} and \emph{p2} stands for \emph{Phase 1} and \emph{Phase 2} respectively.}
  \label{EEG_ablation}
\end{table*}
\begin{table*}[h]
  \centering
  \makebox[\textwidth][c]{\begin{tabular*}{\textwidth}{P{0.16\textwidth}|P{0.1\textwidth}P{0.1\textwidth}P{0.1\textwidth}P{0.1\textwidth}|P{0.08\textwidth}P{0.08\textwidth}P{0.08\textwidth}}
    \hline
    \hline
    \multirow{2}{*}{\textbf{Method}} &
      \multicolumn{4}{c|}{\textbf{BLEU-N (\%)}} &
      \multicolumn{3}{c}{\textbf{ROUGE-1 (\%)}} \\
      &\textbf{BLEU-1}&\textbf{BLEU-2}&\textbf{BLEU-3}&\textbf{BLEU-4}&\textbf{F}&\textbf{P}&\textbf{R}\\
    \hline
    UniCoRN&\textbf{62.90} &	\textbf{49.00} &	\textbf{40.59} &	\textbf{34.77} &	\textbf{59.52} &	\textbf{62.65} 	&\textbf{56.91}\\
    \hline
    w/o p1	&60.74 &	46.02 &	37.27 &	31.25 &	57.41 	&60.69 &	54.69 
\\
    \hline
    w/o p2	&61.91 &	47.36 &	38.66 &	32.66 &	58.33 	&61.36 	&55.78 
\\
    \hline
    w/o p1p2	&53.58 &	35.53 &	25.78 &	19.68 	&48.75 &	53.39 &	45.08 
\\
    \hline
    \hline
  \end{tabular*}}
  \caption{Results of fMRI2text ablation study. \emph{p1} and \emph{p2} stands for \emph{Phase 1} and \emph{Phase 2} respectively.}
  \label{fMRI_ablation}
\end{table*}
Our model utilizes the Pytorch-based \citep{paszke2019pytorch} Huggingface Transformers \citep{wolf2020transformers} packages and is designed to reconstruct sequences with a length of 5 for fMRI2text and 10 for EEG2text in \emph{Phase 2}. Additional hyperparameters can be found in Appendix \ref{appendix-hyperparameter}. Both datasets are split into \emph{train}, \emph{validation}, \emph{test} sets with a ratio of 70\%, 15\%, 15\% respectively. We follow the same evaluation strategy as \citet{wang2022open} to establish a fair comparison and gain insights into the optimal performance scenario of UniCoRN. The results reported are the average of three separate runs. All experiments were conducted on NVIDIA A100-80GB-PCIe GPUs.

\subsection{UniCoRN Structure for fMRI2text}
\paragraph{fMRI2text across Different Splits}
We experiment with series length $\mathcal{T}$ of 10 and report the BLEU scores and ROUGE-1 scores for fMRI2text across different splits. As shown in Table \ref{split_result}, UniCoRN achieves fairly effective results across all splitting methods introduced in Section \ref{dataset}.  Meanwhile, to have an intuitive grasp of the decoding quality, we present a few cases comparing the target tokens and the predicted tokens in Table \ref{case_study_fMRI}.  

The experiments conducted under the \emph{random}, \emph{random time}, and \emph{by subject} settings resulted in BLEU-4 scores of 39.74\%, 34.77\%, and 37.78\% respectively. These results shed light on the prospect of fMRI2text when viewing it as a translation-like task, particularly in comparison to state-of-the-art results in machine translation, such as 46.40\% for English-French translation as reported by \citet{liu2020very} and 15.20\% for English-Arabic translation as reported by \citet{provilkov2019bpe}.

In contrast, the results obtained under the \emph{by stimuli} and \emph{consecutive time} settings are less so ideal. This may be attributed to the fact that the input fMRI frames do not correspond to a fixed and predetermined set of words. Consequently, the fMRI embeddings learned by the model may represent an imprecise combination of words rather than specific, individual words. Such variability might pose a challenge when the model encounters frames paired with unique word combinations unseen during training. Nonetheless, this does not preclude UniCoRN's ability to extract meaningful information under these conditions. As shown in Table \ref{case_study_fMRI}, despite a decline in decoding quality under these two methods, UniCoRN is still successful in identifying key words within the text fragments, and maintains a semblance of polarity and structure that resonates with the target sentence.

One thing to notice is that the results under the \emph{by subject} split setting do not show a significant deviation from those under the \emph{random} and \emph{random time} settings. This contrasts with previous studies that relied on individual fMRI frames for decoding, which suggests that UniCoRN's incorporation of inter-volume information can mitigate the effects of inter-subject variability on decoding performance.

Another interesting anomaly is that, despite that both the \emph{random time} and \emph{consecutive time} configurations have distinct text content across their \emph{train}, \emph{validation}, and cons\emph{test} sets, the former setting performs significantly better than the latter. This discrepancy may be attributed to the robust decoding capabilities of BART, which effectively bridges the gap between frames that UniCoRN did not encounter during training.

The above results demonstrate an intrinsic characteristic when interpreting the fMRI2text task as a translation-like endeavor. The fMRI time series of different subjects can be likened to the unique accent or speaking style that each individual possesses. While variations among individuals exist, they usually do not present significant challenges in discerning the overall meaning, especially when contextual information is provided. This analogy extends to the case of \emph{random time} and \emph{consecutive time}: when a non-native speaker attempts to comprehend a foreign language, the chances of comprehending key information increase significantly when  interpretation can be made from a broader context, as opposed to deciphering a sentence without any foresight of what follows.

\paragraph{Effect of Series Length $\mathcal{T}$}

To further demonstrate the effectiveness of decoding fMRI by series, we conduct experiments on different series length $\mathcal{T}$ under \emph{random time} split setting. As shown in Table \ref{len}, the length of fMRI series does have a major impact on decoding results when $\mathcal{T}$ is relatively small. However, this impact seems to reach a plateau and might even turn adverse as $\mathcal{T}$ increases. Such trend could be attributed to the inherent limitations of the transformer model in effectively learning long-term dependencies.

Meanwhile, although decoding results tend to be less optimal when $\mathcal{T}$ is small, experiments indicate that apart from frequently used phrases (such as catchphrases during pauses), UniCoRN can still decode semantically and syntactically similar tokens. This capability aligns with previous studies, affirming the feasibility of bridging fMRI and human language under naturalistic settings.

\subsection{UniCoRN Structure for EEG2text}

As shown in Table \ref{EEG_ablation}, the UniCoRN structure surpasses the former baseline on all metrics except when solely using snapshot reconstruction, which will be further discussed in Section \ref{ablation}.

Here we take a closer look at the performance of UniCoRN on EEG2text in Table \ref{case_study_EEG} and compare with the former baseline in \citet{wang2022open}. The results illustrate that UniCoRN outperforms the previous baseline in terms of capturing semantics and syntax in target tokens. Specifically, UniCoRN not only enhances the decoding accuracy of individual words but also maintains superior coherence in sentence structure, resulting in more fluent and comprehensible decoding outputs

\section{Ablation Study}\label{ablation}

To further validate the effectiveness of UniCoRN, we conduct ablation studies on both fMRI2text and EEG2text, to assess how the two phases of signal reconstruction affect the model's performance.

As shown in Table \ref{fMRI_ablation}, fMRI2text greatly benefits from both phases of fMRI reconstruction, resulting in an improvement of the BLEU score by approximately 20\% when reconstruction is included. This indicates that for cognitive signals that are rich in spatial information like fMRI, it is important for the encoder to have a thorough understanding of these signals themselves, but not mainly rely on the ability of decoder. Comparatively, series reconstruction proves to be slightly more effective than snapshot reconstruction, which may be attributed to the nature of seq2seq tasks as the input of series reconstruction is more similar to that of cog2text decoding than snapshot reconstruction.

Conversely, Table \ref{EEG_ablation} shows a decline in overall metrics when only \emph{Phase 1} is used for EEG2text. This could be attributed to the noise introduced by the snapshot reconstruction, which might potentially compromise the ability of the model to process EEG sequences — a crucial aspect for cognitive signals with high temporal resolution like EEG. However, this doesn't undermine the importance of snapshot reconstruction for such signals. As evident in the results, combining snapshot and series reconstruction increases the BLEU-4 score from 36.51\% to 37.04\%, suggesting an enhancement in the model's performance for predicting longer n-grams. Thus, while the impact may vary depending on the spatial and temporal resolution of different cognitive signals, integrating both phases generally enhances the model's overall performance by developing a more sophisticated encoder.

\section{Conclusion}

In this paper, we introduce a novel open-vocabulary brain decoding task fMRI2text, aiming to decode linguistic stimuli from multiple fMRI frames collected under naturalistic conditions. Building upon this, we present UniCoRN, a two-stage framework that integrates both temporal and spatial aspects of cognitive signals through snapshot and series reconstruction. The efficacy of UniCoRN is validated under various split settings, illuminating the opportunities that this task provides. Furthermore, we adapt the framework to EEG2text, demonstrating its capacity to generate semantically and syntactically more accurate results, thereby introducing a fresh perspective to brain decoding tasks.
\section*{Limitation} 
The ``Narratives'' dataset provides a valuable fMRI resource, stimulated by language and obtained under naturalistic conditions. Further research opportunities can be pursued with the availability of more detailed datasets. For instance, comparative studies between instances of stuttering and non-stuttering in text stimuli can be conducted, as our experiments demonstrate that the model tends to retain frequently-used filler words (such as ``um'' and ``like,'') as a shortcut for higher accuracy. Meanwhile, the evaluation strategy applied for current research of open-vocabulary brain decoding presents an idealized condition and and serves as a starting point from which further exploration of how existing methods might perform under more real-world scenarios can commence. Although we use this setting for baseline comparison purposes and a testament to the feasibility of our fMRI2text task, additional tests under more practical conditions could be an essential step in future work, further elucidating the applicability and robustness of the methods. Furthermore, the structure of the snapshot encoder can be explored further, as exemplified by the use of transformer-based Vision Transformer (ViT) in \citet{chen2022seeing} for fMRI encoding.
\section*{Ethical Considerations}
In this work, we introduce a new NLP task related to fMRI and a unified approach for decoding various types of cognitive signals into human language. We conduct our experiments on the public cognition datasets \emph{Narratives} and \emph{ZuCo1.0} with the authorization from the respective maintainers of the datasets. All experimental datasets involved have been de-identified by dataset providers and used for research only.
\section*{Acknowledgements}

We express our sincere gratitude to the anonymous reviewers for their professional, insightful and constructive comments and gratefully acknowledge the support of the National Key R\&D Program of China [2021ZD0113302]; and the National Natural Science Foundation of China [62206079]; and Heilongjiang  Provincial Natural Science Foundation of China [YQ2022F006].

\bibliographystyle{acl_natbib}
\bibliography{anthology}
\appendix
\label{appendix}

\section{Implementation Details}
\label{appendix-hyperparameter}
The hyperparameters for the experiments in this paper are shown in Table \ref{hyperparameter}.
\begin{table}[h]
    \renewcommand{\arraystretch}{1.2}
    \resizebox{\columnwidth}{!}{\begin{tabular}{c|c|c|c|c}
    \hline
    \hline
    \multicolumn{2}{c|}{\textbf{Task}}&\textbf{Initial LR}&\textbf{Batch Size}&\textbf{Epoch}
    \\
    \hline
    \multirow{3}{*}{fMRI2text}&p1&1e-3&512&10\\
    \cline{2-5}
    &p2&1e-3&256&5\\
    \cline{2-5}
    &p3&1e-3&224&10\\
    \hline
    \multirow{3}{*}{EEG2text}&p1&1e-4&768&30\\
    \cline{2-5}
    &p2&5e-4&292&30\\
    \cline{2-5}
    &p3&1e-4&16&50\\
    \hline
    \hline
    \end{tabular}}
    \caption{Hyperparameters used in this paper.}
    \label{hyperparameter}
\end{table}

\section{Notation Table}
\label{appendix-notation}
The notation for the variables mentioned in this paper is presented in Table \ref{notation}.
\begin{table*}[h]
    \renewcommand{\arraystretch}{1.2}
    \begin{tabular}{P{0.1\textwidth}|P{0.85\textwidth}}
    \hline
    \hline
    $f_k^{ij}$ & the $k$ th fMRI frame taken when subject $i$ receives stimuli $j$\\
    \hline
    $S_k$ & the subject indexed with k\\
    \hline
    $C_k$ & the stimuli indexed with k \\
    \hline
    $\mathcal{F}^{ij}$ & the collection of all the fMRI frames acquired when subject $i$ receives stimuli $j$\\
    \hline
    $\mathcal{F}^{ij}_{k\sim\mathcal{T}}$ & the fMRI time series of length $\mathcal{T}$ starting at the $k$ th frame\\
    \hline
    $\mathbb{F}_{Tr}$ & the collection of the fMRI time series contained in the training set\\
    \hline
    $T_{Tr}^j$ & the collection of the index of the starting frames of the input fMRI time series from stimuli $j$ in the training set \\
    \hline
    $\mathcal{C}_{Tr}$ & the collection of the index of the stimuli in the training set\\
    \hline
    $\mathcal{S}_{Tr}$ & the collection of the index of the subjects in the training set\\
    \hline
    $E_k^i$ & the snapshot embedding for the $k$ th fMRI frame\\
    \hline
    $E_k^e$ & the serialized embedding for the $k$ th fMRI frame \\
    \hline
    $E^i_{k\sim\mathcal{T}}$ & the snapshot embeddings for fMRI time series of length $\mathcal{T}$ starting at the $k$ th fMRI frame\\
    \hline
    $E^e_{k\sim\mathcal{T}}$ & the serialized embeddings for fMRI time series of length $\mathcal{T}$ starting at the $k$ th fMRI frame\\
    \hline
    $E_{k\sim\mathcal{T}}$ & the fMRI embeddings for fMRI time series of length $\mathcal{T}$ starting at the $k$ th fMRI frame\\
    \hline
    \hline
    \end{tabular}
    \caption{Notations for the main variables used in this paper.}
    \label{notation}
\end{table*}

\section{Details of Dataset}
\label{appendix-dataset}
The detailed information of the ``Narratives'' datasets that are used for fMRI2text experiments in this paper is shown in Table \ref{dataset_fMRI}.
\begin{table*}[h]
    \centering
    \begin{tabular*}{\textwidth}{P{0.45\textwidth}|P{0.1\textwidth}|P{0.1\textwidth}|P{0.1\textwidth}|P{0.1\textwidth}}
    \hline
    \hline
    \textbf{Stimuli}&\textbf{Duration}&\textbf{TRs}&\textbf{Words}&\textbf{Subjects}\\
    \hline
    "Pie Man"&07:02&282&957&82
    \\
    \hline
    "Tunnel Under the World"&25:34&1,023&3,435&23
    \\
    \hline
    "Lucy"&09:02&362&1,607&16
    \\
    \hline
    "Pretty Mouth and Green My Eyes"&11:16&451&1,970&40
    \\
    \hline
    "Milky Way"&06:44&270&1,058&53
    \\
    \hline
    "Slumlord"&15:03&602&2,715&18
    \\
    \hline
    "Reach for the Stars One Smal Step at a Time"&13:45&550&2,629&18
    \\
    \hline
    "It's Not the Fall That Gets You"&09:07&365&1,601&56
    \\
    \hline
    "Merlin"&14:46&591&2,245&36
    \\
    \hline
    "Sherlock"&17:32&702&2,681&36
    \\
    \hline
    "The 21st Year"&55:38&2,226&8,267&25
    \\
    \hline
    Total&  3.1 hours& 7,424 &
    \multicolumn{2}{c}{29,174}
    \\
    \hline
    Total across subjects& 5.0 days& 228,169 &\multicolumn{2}{c}{887,924}
    \\
    \hline
    \hline
    \end{tabular*}
    \caption{Details of the ``Narratives'' dataset used in this paper.}
    \label{dataset_fMRI}
\end{table*}

\section{Details of UniCoRN for EEG2text}
\label{appendix-eeg}
\begin{figure*}[h] 
\centering 
\makebox[\textwidth][c]{%
\includegraphics[width=\textwidth]{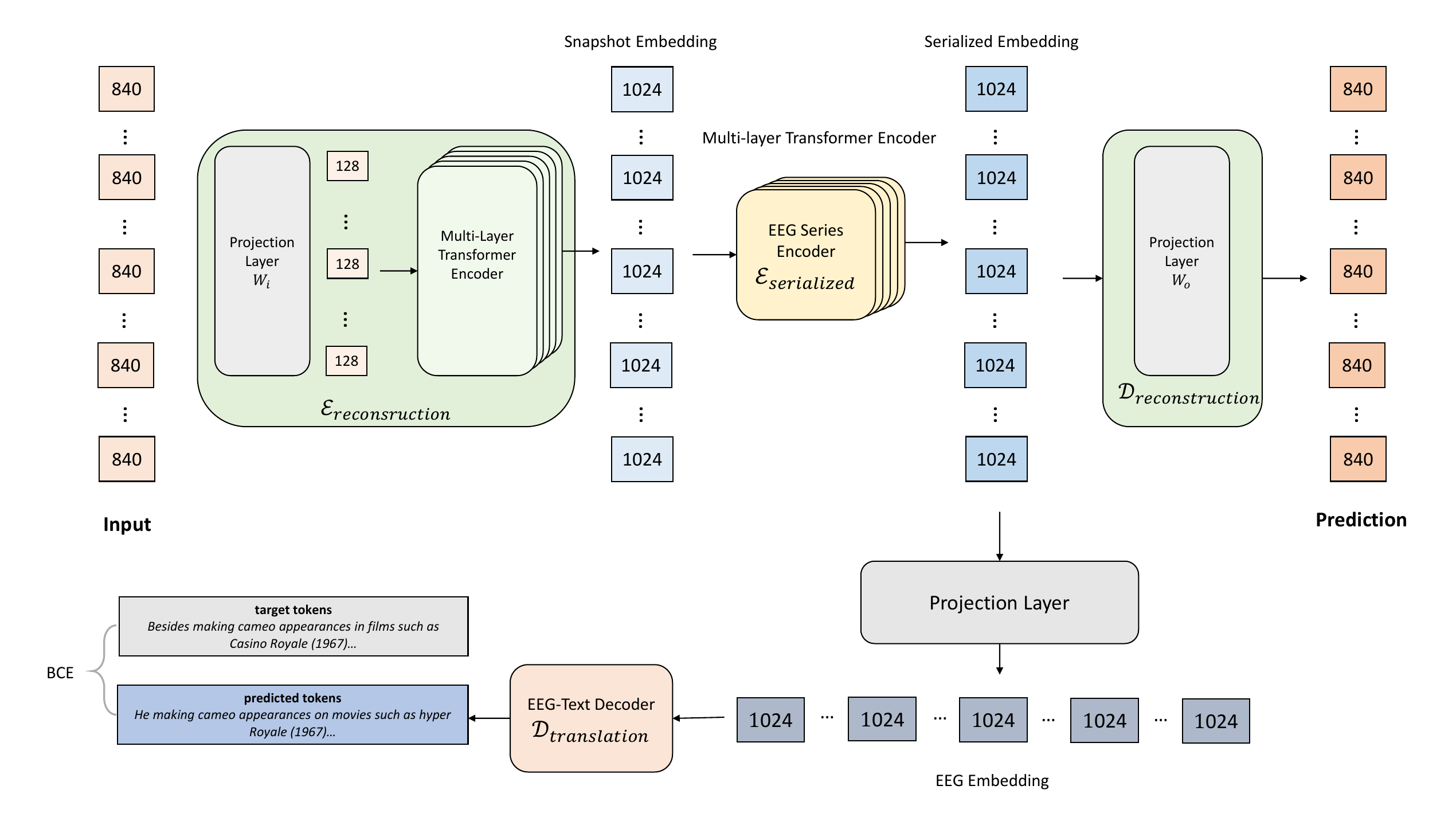}
}
\caption{UniCoRN for EEG-To-Text Decoding} 
\label{eeg} 
\end{figure*}

As depicted in Figure \ref{eeg}, the snapshot encoder $\mathcal{E}_r$ begins by partitioning the original EEG signal into smaller patches. Subsequently, a multi-layer transformer encoder is utilized to analyze the connections between these patches. The resulting output of $\mathcal{E}_r$ is then concatenated and transformed into a vector with a dimensionality of 1024, serving as the snapshot embedding. The subsequent steps in the process are analogous to those used in the fMRI2text scenario.

\section{Case Analysis}
\label{appendix-case}
In this section, we present several cases from our ablation study in Table \ref{ablation_case_fMRI} for fMRI2text and Table \ref{ablation_case_EEG} for EEG2text to provide a more comprehensive understanding of the variations in decoding quality and the impact of different phases.

As demonstrated in Table \ref{ablation_case_fMRI}, UniCoRN effectively decodes ``key information'' ranging from verbs (such as ``swallowing'' and ``smiled'') to nouns (``chocolate'' in this example). Without the series reconstruction in \emph{Phase 2}, the model still demonstrates the ability to decode some nouns, but its performance in predicting verbs is significantly impaired. The performance further deteriorates when the snapshot reconstruction in \emph{Phase 1} is removed, although the model still retains sentence structure that is more similar to the target sentence than the model without \emph{Phase 1} and \emph{Phase 2}.

In contrast, the differences in decoding quality are less pronounced in the case of EEG2text. Although UniCoRN is still able to decode some accurate information such as ``Einstein'' and ``Soviet'', it fails to correctly decode ``physicist'' like other methods, and instead generates ``government''. This discrepancy could be attributed to the fact that EEG signals are aligned at the word level, making the task of decoding EEG less challenging than fMRI2text and thus not showcasing the superiority of UniCoRN as much. Additionally, it could be attributed to UniCoRN's efficient encoder which allows for better utilization of pre-trained language models, since ``government'' might be mentioned more frequently in the context of ``Soviet'' than ``physicist''.
\begin{table*}[h]
    \renewcommand{\arraystretch}{1.6}
    \centering
    \makebox[\textwidth][c]{\begin{tabular*}{\textwidth}{P{0.15\textwidth}|P{0.8\textwidth}}
    \hline
    \hline
    Target Sentence&
    On his way to seat, while \textbf{swallowing} what was left of his \textbf{chocolate}, he \textbf{smiled} to himself.
    \\
    \hline
    UniCoRN& 
    On the way to seat, while \textbf{swallowing} what was left hand his \textbf{chocolate}, Mr \textbf{smiled} to himself.
    \\
    \hline
    w/o p1&
    On his way he get, while \underline{the} what was left with the \underline{lesson}, Mr \underline{was} to be.
    \\
    \hline
    w/o p2&
    What the way to the, while \underline{they} what was left hand his \textbf{chocolate}, he\underline{'d} to himself.
    \\
    \hline
    w/o p1p2&
    and the heart to the, while \underline{she} what was saying hand the mother, Mr \underline{was} to.
    \\
    
    \hline
    \hline
    \end{tabular*}}
    \caption{Case Analysis for fMRI2text ablation study.}
    \label{ablation_case_fMRI}
\end{table*}
\begin{table*}[h]
    \renewcommand{\arraystretch}{1.2}
    \centering
    \makebox[\textwidth][c]{\begin{tabular*}{1\textwidth}{P{0.15\textwidth}|P{0.8\textwidth}}
    \hline
    \hline
    Target Sentence&
    \textbf{Abram Joffe}, a \textbf{Soviet physicist} who knew Einstein, in an obituary of \textbf{Einstein}, wrote...
    \\
    \hline
    UniCoRN& 
    \emph{Heram Joff}, a \textbf{Soviet} \underline{government} who wrote Einstein, in an Americanitken of \textbf{Einstein}, and, wrote...
    \\
    \hline
    w/o p1&
    \emph{Heram J. -}, a \underline{Bachelor} \textbf{physicist} who has Einstein, in a Academyitken of \underline{Win}, was, wasH most...
    \\
    \hline
    w/o p2&
    \emph{Heram Jia (}, a \underline{grades} \textbf{physicist} of is Einstein, and an Americanitken of \underline{his}, and, andB film...
    \\
    \hline
    w/o p1p2&
    \emph{Heram Joff about}, a \underline{family} \textbf{physicist} who was Einstein, in an Americanitken of \underline{an}, in, NewC...
    \\
    
    \hline
    \hline
    \end{tabular*}}
    \caption{Case Analysis for EEG2text ablation study.}
    \label{ablation_case_EEG}
\end{table*}

\clearpage

\end{document}